# Demonstration of Achromatic Cold-Neutron Microscope Utilizing Axisymmetric Focusing Mirrors

D. Liu[1], D. Hussey[2], M.V. Gubarev[3], B. D. Ramsey[3], D. Jacobson[2], M. Arif[2], D. E. Moncton[1,4], and B. Khaykovich[1a]

[1]Nuclear Reactor Laboratory, Massachusetts Institute of Technology, 138 Albany St., Cambridge, MA 02139, USA

[2]Physical Measurement Laboratory, NIST, Gaithersburg, MD, 20899-8461 USA

[3]Marshall Space Flight Center, NASA, VP62, Huntsville, AL 35812, USA

[4]Department of Physics, Massachusetts Institute of Technology, 77 Massachusetts Ave., Cambridge, MA 02139, USA

An achromatic cold-neutron microscope with magnification 4 is demonstrated. The image-forming optics is composed of nested coaxial mirrors of full figures of revolution, so-called Wolter optics. The spatial resolution, field of view, and depth of focus are measured and found consistent with ray-tracing simulations. Methods of increasing the resolution and magnification are discussed, as well as the scientific case for the neutron microscope. In contrast to traditional pinhole-camera neutron imaging, the resolution of the microscope is determined by the mirrors rather than by the collimation of the beam, leading to possible dramatic improvements in the signal rate and resolution.

Modern optical instruments for visible and synchrotron light use a variety of focusing devices, such as lenses, Fresnel zone plates and mirrors. These devices help increase the signal rate, resolution, or both. Were such powerful optical tools available for neutron scattering, they might bring significant, even transformative, improvements to rate-limited neutron methods and

---

a) Electronic mail: bkh@mit.edu

enable new science. We have recently demonstrated such a tool: grazing-incidence mirrors based on full figures of revolution, often referred to as Wolter mirrors.[1,2] These mirrors have the potential of transforming neutron imaging (and scattering) instruments from pinhole cameras into microscopes.

Currently, the most commonly used optical devices are neutron guides, which transport neutrons, but do not create images.[3-7] Imaging optics, such as lenses and mirrors, can be found at some reactor-based small-angle scattering (SANS) instruments.[8-12] Imaging with the help of lenses has been demonstrated,[13] but not used due to practical limitations. Lenses are not suitable for time-of-flight (TOF) instruments, which use polychromatic beams, since the focal distance of a biconcave neutron lens varies inversely as the second power of the neutron wavelength. A recently developed magnetic lens offers a possible solution, at least for very cold neutrons,[14,15] by modulating the magnetic field according to the neutron velocity as a pulse travels through the lens. However, these lenses work only with polarized pulsed neutron beams.[16,17] In contrast to lenses, focusing mirrors are free from chromatic aberrations. Flat mirrors can be bent to create Kirkpatrick-Baez[18] optics and other combinations of parabolic and elliptical mirrors,[19] but these systems have very limited collection efficiency for large, divergent neutron beams. As a result, only one SANS instrument is equipped with toroidal Cu-coated mirror,[20] but no imaging instrument uses reflection optics (except for demonstrations of a microscope based on ultra-cold neutrons,[21,22] but such devices are not practical for modern neutron radiography). The limitation on the collection efficiency arises from the exceedingly small critical incidence angle, above which almost no neutrons are reflected. To work around this limitation and dramatically improve the performance of mirror-based instruments, we have used optical designs and technology inspired by x-ray telescopes.[1,23-27] This paper reports the demonstration of a prototype cold-neutron microscope, which is equipped with image-forming Wolter mirrors. Images of test samples, such as those in Figure 1, were taken and analyzed. The resolution, throughput, depth of focus, and field of view were found in agreement with ray-tracing simulations.

Neutron imaging is a valuable non-destructive evaluation method for many applications, such as multi-phase flow in porous media, water transport in fuel cells, degradation mechanisms in lithium batteries, car engines, and cultural heritage objects.[28] Traditional neutron-imaging instruments resemble pinhole cameras, as shown schematically in Figure 2(a). The spatial resolution is limited largely by the beam collimation, which is characterized by the L/D ratio, where D is the source aperture through which samples are illuminated, and L is the aperture-to-object distance. To achieve high L/D for a reasonable instrument length, D is often less than 10 mm, severely restricting the flux illuminating the sample. The goal of the imaging community is to achieve the spatial resolution of ~1 µm, without reducing the signal rate to impractically low levels. Available neutron fluxes and detector resolution have so far prevented reaching this goal, but the focusing optics demonstrated here can be used to overcome these limitations.

Focusing mirrors make the optical design of a neutron-imaging instrument similar to that of optical microscopes (see Figure 2(b)), where the resolution does not depend on the beam collimation, but rather on the optics itself (since the cold neutron wavelength is ~1 nm, diffraction does not limit current resolution). Consequently, the source size, and thus the number of neutrons illuminating the sample increase substantially, leading to higher signal rates. In addition, optical magnification would result in a better spatial resolution at the same pixel size of the detector. Therefore, the use of Wolter optics opens the possibility for significant progress in high-resolution thermal neutron radiography, similar to the use of lenses in optical devices.

Schematic illustrations of two experimental set-ups used at the NIST Center for Neutron Research (beamline NG1) are shown in Figure 2. The mirrors used for the measurements are described in Ref. [1]. This prototype device includes three nested coaxial mirrors, each one consisting of confocal ellipsoid and hyperboloid pairs. The optic was placed in the beamline with 4 degrees of motorized control (pitch, yaw, longitudinal and horizontal transverse). The samples and the detector were at the foci of the optics (0.64 m upstream and 2.56 m downstream, respectively). Two test samples were used to measure the spatial resolution, depth of field and figure errors. One was a planar grating (referred to here as G0) made of about 5 μm–thick neutron-absorbing Gd,[29] with a period of 796 μm and a duty cycle of 40 %, and the other was a planar array of 100 μm-diameter holes in a 100 μm-thick Gd foil (a modified uniformly redundant array, or MURA).[30] The sample position was about 2 m from the end of the guide (cross sectional area 60 x 60 mm). There was a 50 mm-diameter shutter between the guide and the sample. The detector is a 40 mm-diameter micro-channel plate (MCP) with a cross-strip readout. At full resolution, the detector is a 64 mega-pixel camera, with 4.98 μm pixel pitch.[31]

Figure 1 shows images of the samples obtained with and without the optics. In the conventional system without the optics, the source-to-detector distance of 3.2 m and the aperture diameter of 3 mm were used; thus, the L/D ratio was about 1000, as normally used for high-resolution imaging. When using the mirrors, the source aperture was enlarged from 3 mm to about 10 mm. The average neutron flux at the detector increased from 7 to 20 counts per pixel when the conventional system was replaced by the microscope. Such increase, by a factor of three, is consistent with ray-tracing simulations. From acquired magnified images, one can derive the modulation transfer function (MTF), which is commonly used to quantify the spatial-frequency response of optical systems.[32] (MTF decreases from unity with increasing frequencies, because the inevitable blurring induced by the imaging system reduces the contrast of high-spatial-frequency components of images. When the contrast is too small, high-frequency features cannot be resolved.) Normally, the resolution of a neutron imaging system is derived from MTF = 0.1. Figure 3 shows the measured and simulated MTF of the neutron microscope, derived from cross-sections such as this shown in Figure 1(c). (Fourier transforms of the first derivatives of edge-spread functions were calculated, according to the standard procedure.[32]) MTF = 0.1 corresponds to 0.0035 μm$^{-1}$. Considering the magnification M = 4, the spatial resolution is about 70 μm, equivalent to 110 μrad angular resolution (since the focal length of the optics is 0.64 m), in complete agreement with x-ray measurements, which were made after fabrication of the mirrors. The traditional set-up had a resolution of 115 μm.

The spatial resolution of a mirror-based imaging system is determined mainly by deviations of the actual mirror surface from its ideal shape. Effects of long-wavelength deviations, called figure errors (or slope errors), can be taken into account by geometrical optics. Therefore, these effects were simulated by ray-tracing. We did not simulate weaker effects of short-wavelength deviations, or surface roughness, which lead to off-specular scattering due to diffraction and incoherent scattering and cannot be simulated by ray-tracing alone. Ray-tracing simulations of the experimental setup were performed using McStas, a standard neutron ray-tracing package.[33,34] Because of figure errors, neutrons are reflected into a range of angles around ideal specular reflections. Therefore, in the simulations, scattering angles were assigned random deviations from ideal specular directions, according to a two-dimensional Gaussian distribution with standard deviation σ. Different standard deviations resulted in different images, from which MTF's were extracted. Figure 3 shows such MTF curves corresponding to standard deviations σ = 40 μrad, 20 μrad and 1 μrad, together with points deduced from the experimental image. The value of σ = 20 μrad in simulations leads to a similar spatial resolution as that measured. Similar slope errors were found by x-ray testing, following mirror fabrication. We attribute the low-spatial-frequency deviation of the measured and simulated curves to the effects of roughness, which leads to off-specular scattering at large angles. It is responsible for a weakly changing background that manifests itself mostly at low spatial frequencies, while figure errors influence strongly the high-spatial-frequencies part of the MTF. Effects of the surface roughness will be taken into account in future simulations.

The field of view (FOV) and depth of focus (DOF) were characterized with the help of the thin Gd MURA mask (Figure 1(c)). A larger DOF means sharper images of three-dimensional objects and higher tolerance to the displacements along the optical axis from the focal plane. The DOF was measured by scanning the sample along the optical axis near the nominal focal point. We analyzed fragments of images, containing three adjacent pinholes (shown in the inset in Figure 4). The horizontal profile across the pinholes was fitted with Gaussians and the full width at half maximum (FWHM) of each peak was extracted. Figure 4 shows the average FWHM vs. positions of the sample along the beam axis. The DOF, where FWHM is approximately constant, is about 6.5 mm independent of the position across the FOV. Because of the four-fold magnification and the 40 mm diameter MCP detector, the FOV has a diameter of 10 mm. By ray tracing, we found the FOV to be about 12 mm. To evaluate the image quality across the FOV, we examined different triplets of adjacent apertures along vertical and horizontal directions in the image plane. The FWHM was found constant along both directions, thus the image quality was uniform everywhere in the FOV.

In a conventional pinhole imaging system, the spatial resolution increases approximately linearly with the L/D ratio, while the neutron flux at the sample increases approximately as $(D/L)^2$. Therefore, high spatial resolution requires significant losses of neutron flux, and *vice versa*. In contrast, the resolution of a neutron microscope is not determined by the L/D ratio. Therefore, when using the mirrors, the sample can be illuminated by a much larger neutron source. The signal rate is improved by increasing both the effective source area and the solid angle of the collected beam. When the optics is well optimized, including the use of nested coaxial confocal mirrors and high-critical-angle coating,[23] enhancements of orders of magnitude in signal rates could be possible.[1,35,36] In addition to the signal rate, the field of view[37,38] and depth of focus also require optimization of the mirrors' geometry.

The spatial resolution of neutron imaging is currently limited by the detector pixel size and available neutron fluxes. The resolution can be improved with the help of magnifying Wolter optics, which can practically be made with up to ten-fold magnification. For example, using standard micro-channel-plate detectors with 5 μm pixels and magnification-ten optics, it could be possible to reach nearly 1 μm resolution in neutron radiography, potentially enabling new science. To achieve this, the resolution of the optics, currently of order 100 μrad, must be improved by a factor of ~100, by reducing figure errors and surface roughness. One technology, differential deposition, is currently being developed for grazing-incidence x-ray telescopes,[39] and could be applied to neutron mirrors to yield the needed surface finish of 1 μrad. (For comparison, the Chandra x-ray telescope, which consist of 4 nested pairs of parabolic and elliptical mirrors, has angular resolution ~0.1 μrad).

The use of large-magnification high-resolution optics would help neutron imaging reach 1 μm length-scales with reasonable image acquisition times. An estimate of the increased signal rate for a 1:1 optic illustrates the potential gains. At the NIST neutron imaging facility, slit apertures of 2 mm x 20 mm are used to obtain images of fuel cells with about 10 μm spatial resolution. The short dimension of the slit provides high resolution along the through-plane of the fuel cell, while the long dimension provides a higher neutron fluence rate, which reaches about $2 \times 10^6$ cm$^{-2}$ s$^{-1}$. This low fluence rate requires an image acquisition time of about 10 min for an accurate measurement of the water content in a fuel cell.  Our estimates for the throughput of the Wolter optics suggest that the effective neutron fluence rate could reach $2 \times 10^8$ cm$^{-2}$ s$^{-1}$ for a gain of 100 over the pinhole optics geometry. Further, the resolution of the image is the same in all directions, as opposed to being blurred, as is the case in the slit image. With such intensity gains it will possible to obtain images with high temporal (~5 s) and spatial resolution (~10 μm) of transient processes in fuel cells, batteries, and microfluidic devices.  As well, tomography data sets with 10 μm voxel resolution could be obtained in a period of less than an hour as opposed to the current ~1 day, and would enable studies of crack formation in concrete during freeze/thaw cycles or slow fluid migration in soils, clays and shale.  The possibility of magnification 10, coupled with a detector spatial resolution of ~10 μm means that the system resolution will reach ~1 μm.  Since the neutron density will be decreased by a factor of 100 for magnification, image acquisition times will be comparable to the current image acquisition time for ~10 μm resolution images.  With this gain in resolution, one would be able to image the water content in standard Pt/C electrodes in fuel cells, and high resolution maps of the lithium concentration in lithium batteries.  Another advantage of an optic for imaging is more difficult to quantify; the sample no longer needs to be in direct contact with the detector and there is ample space both upstream and downstream of the sample.  In studies of magnetic materials with polarized neutrons, the measurement would not suffer a loss in spatial resolution due to the thickness of the neutron spin analyzer.  One could also perform simultaneous x-ray and neutron micro-tomography of porous media and in one unique instrument obtain the complementary information from both probes; x-rays to measure the matrix and neutrons to measure the fluid.

In conclusion, we have demonstrated the first grazing-incidence neutron microscope and analyzed its performance. Measurements of spatial resolution, depth of focus and field of view of the microscope showed a good agreement with ray-tracing simulations. Thus, we have confidence that new mirrors, designed and optimized for a future imaging users facility, would perform as expected. With optimized mirrors and beamline configurations, the improvement of neutron flux and resolution could be very significant, allowing game-changing improvements of the neutron-imaging technique and leading to new science.

**Acknowledgements.**

The authors are grateful to the NCNR technical support staff, especially Eli Baltic and Danny Ogg and to R.G. Downing for the temporary use of NG-1 for these experiments. Research supported by the U.S. Department of Energy, Office of Basic Energy Sciences, Division of Materials Sciences and Engineering under Award # DE-FG02-09ER46556, # DE-FG02-09ER46557. NIST authors acknowledge support from the U.S. Department of Commerce, the NIST Radiation and Biomolecular Physics Division, the Director's office of NIST, the NIST Center for Neutron Research, and the Department of Energy interagency agreement No. DE_AI01-01EE50660.

**FIGURES**

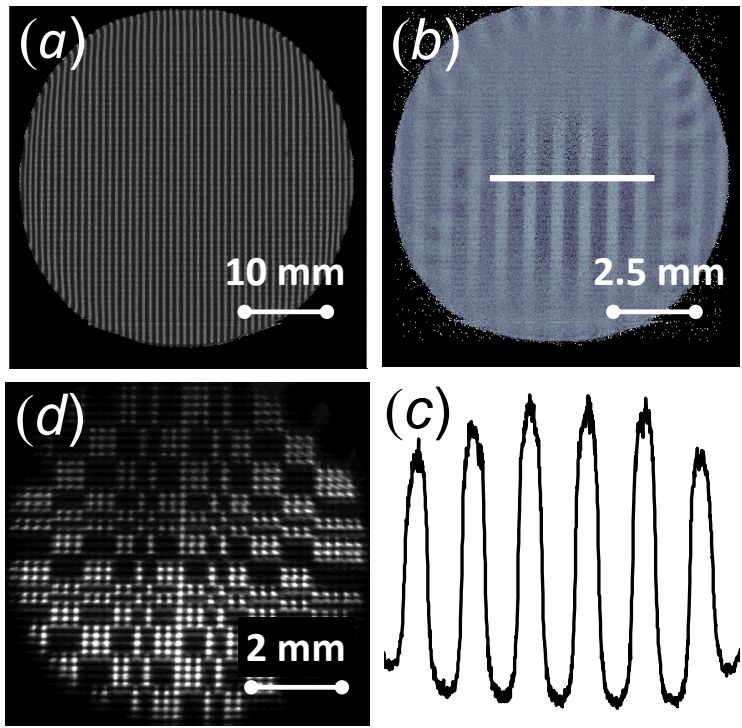

**Figure 1**. Images of test samples made with the pinhole set-up and focusing-mirrors microscope. The samples are: a planar grating (G0) made of about 5 μm–thick neutron-absorbing Gd deposited on quartz with a period of 796 μm and a duty cycle of 40 % (see Ref. 29), and the array of 100-μm-diameter holes, generated from modified uniformly redundant array (MURA), see Ref. 30. (a) Pinhole image of G0; 5 min measurement time. (b) Magnified image of G0; 5 min measurement time. The white line shows the region where the cross-section was taken. (c) Cross-section of the image in (b), taken along the white line. (d) Magnified image of MURA; 15 min measurement time. The inhomogeneity in illumination in the microscope images results from the 2-m-long separation between the end of the neutron guide and the sample. Therefore, the beam is excessively collimated, such that some of the neutrons illuminating the samples do not reach the reflecting mirrors. This effect is avoided when the beam has enough divergence, for example, by placing samples closer to the guide. The one-dimensional cross-section of the MURA image is in the inset of Figure 4.

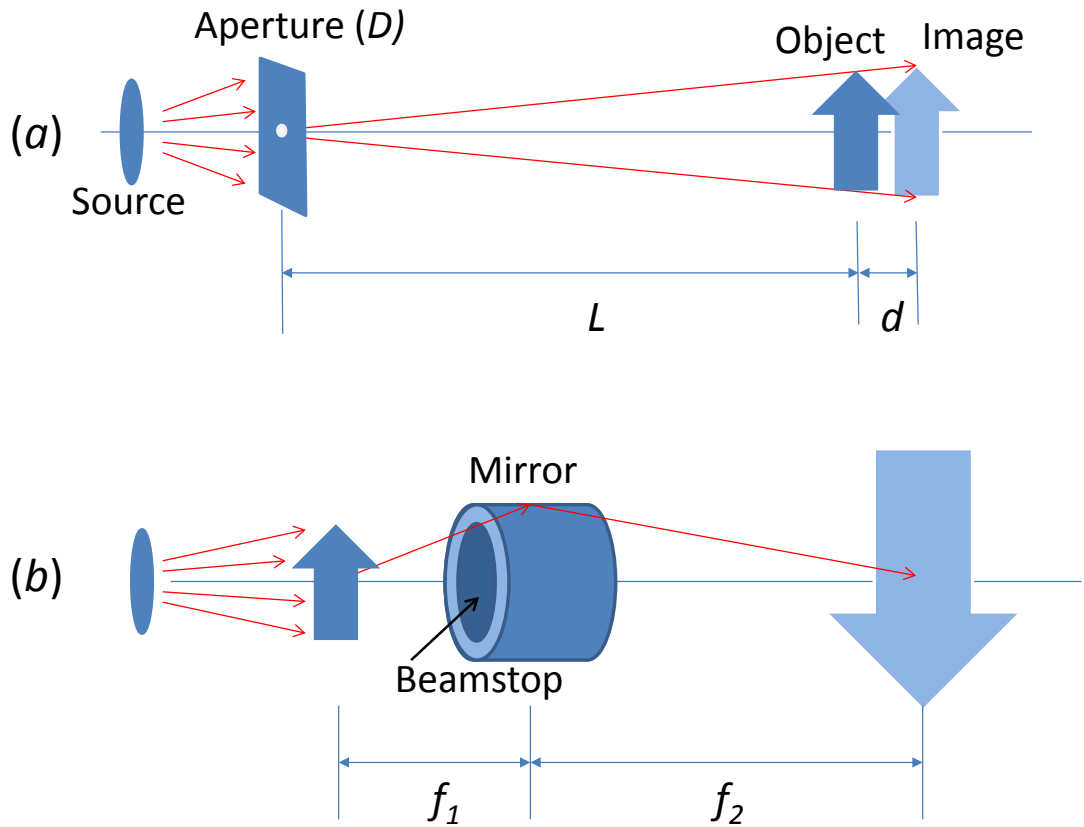

**Figure 2**. Schematic illustrations of neutron imaging instruments. A conventional pinhole-imaging system using a small aperture (a) and a microscope, equipped with axisymmetric grazing-incidence mirrors (b). The mirrors used in this paper have diameters of approximately 30 mm and the total length of 60 mm. The focal distances are $f_1$ = 0.64 m and $f_2$ = 2.56 m, the source to detector distance is 3.2 m and the magnification is 4.

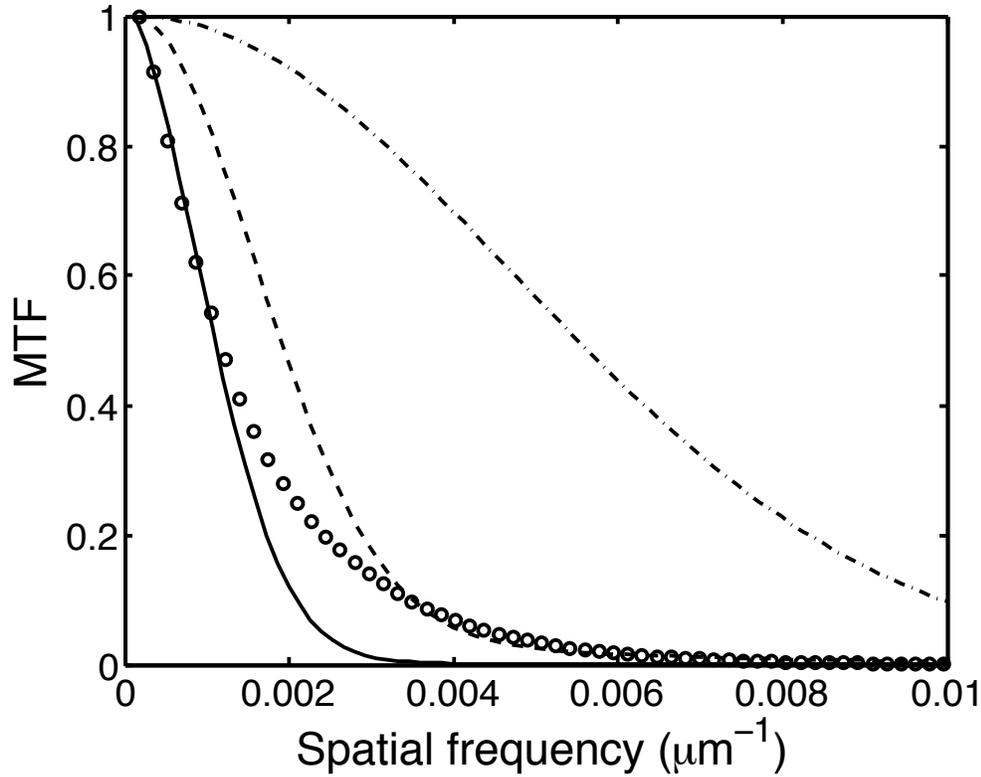

**Figure 3.** MTF analysis using an image of the G0 sample. Open circles are computed by the Fourier transform of the Gaussian-Lorentzian fit of the measured line-spread function (see Figure 1(c)). The spatial frequency at 10% MTF is 0.0035 μm$^{-1}$ corresponding to the spatial resolution of 70 μm. The simulated curves were produced by ray-tracing, using the Gaussian distribution of slope errors with standard deviations σ = 40, 20, and 1 μrad, which are represented by the solid line, the dashed line and the dot-dashed line respectively.

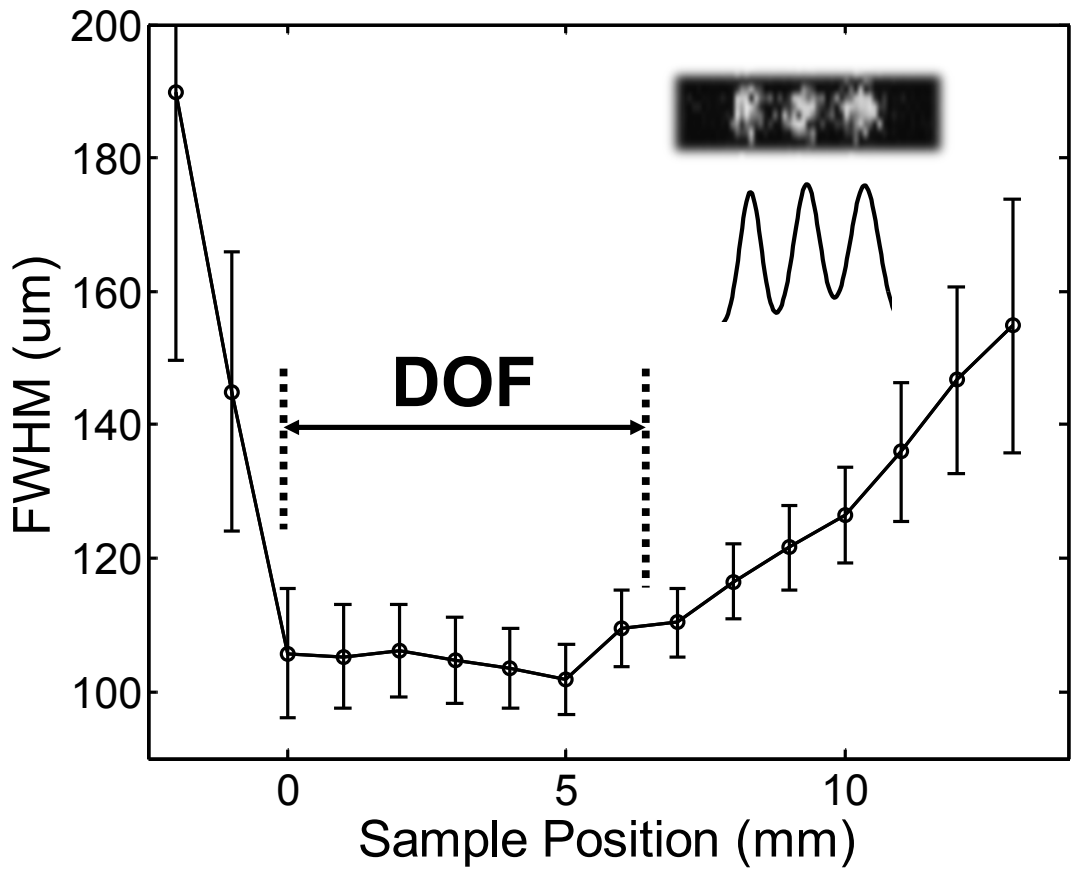

**Figure 4.** Depth of focus analysis. The inset shows the image of a group of three pinholes used for the analysis, and the horizontal profile of the intensity. X-axis represents the position of the MURA mask along the optical axis, with zero at the nominal focus. Y-axis is the average FWHM of the three peaks. The inset was measured at the sample position of 3 mm. The peak widths are approximately constant within the margin of error of about 5 μm; variations of the width are consistent with size variations of the holes themselves due to manufacturing errors. Error bars for the FWHM plot were calculated by fitting the peaks with Gaussian. The depth of focus, where the FWHM is constant within error bars, is about 6.5 mm, as indicated by two dashed lines.